# An Endogenous Mechanism of Business Cycles

Dimitri Kroujiline*[1], Maxim Gusev[1], Dmitry Ushanov[1], Sergey V. Sharov[2], Boris Govorkov[3]

## Abstract

This paper suggests that business cycles may be a manifestation of coupled real economy and stock market dynamics and describes a mechanism that can generate economic fluctuations consistent with observed business cycles. To this end, we seek to incorporate into the macroeconomic framework a dynamic stock market model based on opinion interactions (Gusev et al., 2015). We derive this model from microfoundations, provide its empirical verification, demonstrate that it contains the efficient market as a particular regime and establish a link through which macroeconomic models can be attached for the study of real economy and stock market interaction. To examine key effects, we link it with a simple macroeconomic model (Blanchard, 1981). The coupled system generates nontrivial endogenous dynamics, which exhibit deterministic and stochastic features, producing quasiperiodic fluctuations (business cycles). We also inspect this system's behavior in the phase space. The real economy and the stock market coevolve dynamically along the path governed by a stochastically-forced dynamical system with two stable equilibria, one where the economy expands and the other where it contracts, resulting in business cycles identified as the coherence resonance phenomenon. Thus, the incorporation of stock market dynamics into the macroeconomic framework, as presented here, allows the derivation of realistic behaviors in a tractable setting.

Keywords: Business cycles, coupled real economy and stock market, macroeconomic instability, expectations dynamics, interactions-based models, news-price-news feedback, stochastically-driven dynamical systems, coherence resonance.

[1] LGT Capital Partners, Pfäffikon, Switzerland. [2] Lobachevsky State University, Advanced School of General and Applied Physics, N. Novgorod, Russia. [3] Multiscale Trading Systems, Euthal, Switzerland. *Contact author: dimitri.kroujiline@lgt.com.

# I

It is widely accepted that business cycles emerge due to exogenous shocks stochastically forcing the economy subject to various frictions. Such business cycle models, referred to as DSGE, have been shown to replicate reasonably well the actual dynamics in a relatively broad range of economic regimes. The DSGE models, however, face challenges explaining large fluctuations such as the Great Recession. And although it might be possible to remedy their certain shortcomings by extending standard models (Linde et al., 2016; Christiano et al., 2018), there remains conceptual dissatisfaction regarding the lack of a sound justification for multiple persistent orthogonal shocks that must force the DSGE economy to fit empirical data (Kocherlakota, 2010; Romer, 2016; Blanchard, 2016; Korinek, 2017; Stiglitz, 2018).

This paper argues that the mechanism underlying business cycles may be better understood in a macroeconomic framework that includes out-of-equilibrium stock market dynamics. Our objective here is to examine whether the incorporation of a specific dynamic stock market model, based on opinion interactions, may generate business cycles without support from structural shocks. In particular, this paper contributes to the ongoing discussion by identifying and explaining an endogenous mechanism capable of generating economic fluctuations consistent with observed business cycles.

To this end, we consider a news-driven stock market model developed in Gusev et al. (2015). This model, based on micro-level interactions among agents, yields at a macro level a closed-form nonlinear dynamical system that explains price formation as stemming from endogenous dynamics between information flow, investor expectation and the market price itself.

We begin by deriving this model. This derivation differs from the original as we take a more tractable albeit less formal approach. We then report the results of the empirical tests to validate this model. In addition, we show that the model encompasses the efficient market's random walk as a special case – supporting the claim of the model's generality.

Next, we integrate this stock market model into a macroeconomic system. Our main objective is to ascertain the potential of this approach, so that we limit the scope of research to the leading-order effects. To illustrate these effects, we attach the model to a very simple economic relation that dynamically links the economy's output with aggregate demand while emphasizing the influence of stock market price on spending (Blanchard, 1981).

We then proceed to investigate this coupled real economy – stock market system. Despite the simple economic relation applied, the system exhibits realistic dynamics that contain several time-scales of interest, ranging from daily market price changes to the interaction between the economy and the market over very long time horizons (e.g. decades); the latter, as will be shown, may act as a "transmission mechanism" between technological progress and the long-term economic growth.

The system's tractability enables us to study in detail the endogenous mechanism behind the output fluctuations at business cycle frequencies. This mechanism contains both stochastic and deterministic features. In summary, the variation in the economy's growth modulates the news flow (e.g. company earnings, industry outlooks, economic indicators) over a business cycle, amplifying the positive (negative) market dynamic during the economy's expansion (contraction) as the volume of good (bad) economic news increases. This reinforcing influence of the economy on the market diminishes at the later stages of expansion and contraction. As a result, market price becomes more sensitive to regime-contradicting news, which makes it easier for a random news event to initiate the fast dynamic of market crash or rally that propels the economy, respectively, from expansion to contraction or vice versa. In accordance with this mechanism, it appears that to the leading order the economy determines the business cycle length whereas the market determines its amplitude.

We also inspect the system's behavior in the phase space. The system has a single unstable- and two stable equilibrium points. One stable equilibrium corresponds to the bull market and the expanding economy, while the other describes the bear market and the contracting economy. The stable equilibria are acting as attractors that tend to entrap the economy in the contraction or expansion regime. This entrapment is asymmetric as the influence of technological progress causes the system to stay on average longer in the expansion regime, resulting in long-term economic growth. In the meantime, the flow of exogenous news, through its impact on market price, perturbs the system stochastically and does not allow it to settle at the stable equilibria. In addition, the news flow may occasionally force the system across the regions of different dynamics – thereby triggering regime transitions.

Thus, according to our model, the stock market and the real economy make up a coupled nonlinear system that evolves dynamically under the action of the endogenous forces pulling it toward one of the two equilibrium states and the stochastic force exerted by exogenous news exciting the system and triggering regime changes. We show that these dynamics produce quasiperiodic fluctuations in output with a distinct maximum of the frequency distribution. Using the phase space description, we identify this business cycle mechanism – outlined above in terms of the economy-market interaction – as a particular type of coherence resonance, a relatively recently reported effect in stochastically-forced dynamical systems (Pikovsky and Kurths, 1997).

## II

Since a particular stock market model is at the heart of this study, it may be helpful to outline the relevant conceptual frameworks as well as the recent advancements with respect to stock market modeling in order to provide a proper context for the present work. In addition, we make use of this discussion to lay out the principles behind the model applied and explain its differences from the models proposed in the previous literature.

To model the stock market from microfoundations, it is necessary to postulate how market participants make investment decisions using the available information and, from these assumptions, derive their aggregate behavior influencing price changes. Investment decisions, like any other decision, can be made independently or by following somebody else's ideas. Because investors are generally free to employ either approach, investment decision-making is likely in aggregate to involve both the independent thinking and the external influences.

The classic academic approach assumed non-interacting rational investors, thereby neglecting the impact of external influences on decision-making. This approximation yields the efficient market model with no intrinsic dynamics, where market price follows a random walk driven by stochastic informational shocks on the back of the long-term growth due to technological progress and inflation. This model has been instrumental both as a theoretical framework for understanding market behaviors and in terms of practical application, advancing index investing and other innovations.

Still, the efficient market is a construct built upon relatively heavy assumptions. As such, there are limits to its applicability: it cannot explain frequent large price movements while the em-pirical evidence for certain answers provided by it, such as those related to return unpredictability[1], is inconclusive or contradicting. To account for these discrepancies, theoretical literature in finance has sought to extend the classic paradigm.

Thus, Campbell and Cochrane (1999) and Bansal and Yaron (2004) added time-varying consumption; Sheinkman and Xiong (2003) considered investors with heterogeneous views on asset fundamentals; Gabaix (2012) and Wachter (2013) introduced a time-varying probability of economic disasters; and Barberis et al. (2015) included investors with return expectations extrapolated from past performance – to name but a few among the recent papers that incorporate additional features into the traditional framework in order to obtain various dynamic behaviors.

The present paper belongs to another strand of the literature that aims to improve the explanatory capacity of asset pricing models. To map out the boundaries of the efficient market applicability, this branch of research replaces the crucial assumption of no interaction among investors by the opposite limiting case. That is, it assumes that investors, or some groups of investors, make decisions through interaction alone by exchanging opinions with other investors and/or by accessing opinions expressed in the disseminated information. This approach relies, sensibly, on a premise that it may be less costly for an investor to take an expert opinion than to form an opinion independently. Indeed, anecdotally, market practitioner decision-making is largely based on the opinions that propagate across the investment community via interaction and discourse.

---

[1] A large literature has examined the predictive power of economic variables, such as price/dividend and other valuation ratios. See, for example, a short survey in Section 4 of Campbell (2014).

This research direction seems promising because once opinion interaction has been activated, the link between the micro- and macro properties ceases to be trivial. As a result, aggregate behavior can acquire features not present at the micro level where individual interactions occur. For example, the collective effects emerging through interaction can generate instabilities, known as phase transitions, triggering a nonlinear largescale response to small perturbations. Therefore, this class of models has a potential for describing actual market behaviors, which include bubbles, crashes and trends, more accurately than the models with non-interacting investors. It also helps that the mathematical methods relevant for investigating the interactions-based models have already been developed in statistical mechanics.

A number of such models have been proposed.[2] They seek to generate market dynamics primarily through individual interactions among heterogeneous agents pursuing different investment strategies or beliefs, such as the noise traders (herein, used as a collective term for "irrational" investors) and the fundamental traders, ubiquitous in the market modeling literature.[3] Over time these

---

[2] See, for example, reviews by Brock and Durlauf (2001) and Lux (2009).

[3] Some of the first works that studied price distortions in heterogeneous market models populated by the (non-interacting with respect to opinion exchange) noise- and fundamental traders were Zeeman (1974), Beja and Goldman (1980), Day and Huang (1990), De Long et al. (1990). Interactions-based models that adopted this approach added intrinsic nonlinearities to price dynamics – as nonlinear effects arise naturally via opinion interactions – enriching model behaviors (e.g. Lux, 1995; Lux and Marchesi, 1999). Other interactions-based models studied heterogeneous agents with different investment horizons and extrapolative expectations (Levy et al., 2000), examined the effects of investor clustering (Cont and Bouchaud, 2000) or attempted to explain market regime switches as phase transitions (Vaga, 1990), to name a few among various approaches. The

models have evolved toward incorporating increasingly more realistic assumptions, capable of simulating complex and diverse networks of interacting agents.[4] This achievement came, however, at the cost of a loss in the models' tractability, exacerbated by the sensitivity to the choices made at the micro level, whereas it is often difficult to economically justify any one particular choice over another. Most importantly, although these models can, as expected, simulate certain distinctive features unexplained by the efficient market, notably the non-normal return distribution, they have to the best of our knowledge failed to accurately replicate the market price track as well as to predict in any practical sense market returns. Given these shortcomings, the heterogeneity of trading strategies can hardly be the primary source of the real-world market dynamics.

So, what is then the main source of market dynamics and can it be captured by an agent-based market model or not? Let us suggest the following argument. To acquire internal dynamics, any system must have some mechanism that couples its key variables. In the market, it can be argued, information and price are mutually coupled via a feedback mechanism whereby news cause price changes while price changes draw media response triggering further news releases, which in turn incite further price changes and so forth. Information does not of course impact price directly. It influences expectations that lead to investment decisions and it is these decisions aggregated across the investment community that finally impact market price. It follows that the feedback-interlinked

---

common feature of these models is that being inherently nonlinear they look for explanations of market behaviors as nonlinear phenomena. We also take note of Franke (2014) who proposed a generic model for investor opinion dynamics, which exhibits certain behaviors similar to those found in Gusev et al. (2015).

[4] See, for example, a survey of computational agent-based market models by LeBaron (2006).

information, expectation and price may be the relevant macro variables for describing market dynamics, and thus with the correct choice of hypotheses about how interactions occur at the micro level a dynamic model for the evolution of these variables can be developed that may be able to explain the observed market behaviors.

Gusev et al. (2015) developed, and Kroujiline et al. (2016) extended, such a dynamic model. It is formulated on a micro level as an agent-based model with two types of interacting agents: investors who trade according to their expectations and analysts who interpret news, form expectations and channel them to investors. This formulation yields on a macro level a closed-form dynamical system that explains market behaviors in terms of nonlinear interaction between information flow, investor expectation and market price.[5]

We apply this model to study the economy-market interaction for three reasons. First, it can be easily embedded, through information flow, into macroeconomic models as we will see in Section 2. Second, the model equations are in analytic form, which facilitates the interpretation of combined effects. Third, and more importantly, this model is supported by empirical tests: it can replicate past prices within reasonable tolerance and predict returns with a precision sufficient for designing a successful trading strategy (Kroujiline et al., 2016).

---

[5] The idea that the observations of price changes generate feedback that may significantly affect market dynamics is not new (see a review by Shiller, 2003) and, indeed, models based on interactions usually apply some form of price feedback. However, Gusev et al. (2015) was the first work that presented a micro-level description of the news-price-news feedback mechanism and developed an interactions-based model micro-founded in this mechanism.

III

This paper is organized as follows. Section 1 inspects the dynamic market model described above. We derive this model, then discuss the results of empirical validation and finally determine a set of assumptions under which the model describes the efficient market. Section 2 derives the equations of the coupled economy-market system. Section 3 is a study of this system. We determine the realistic values of parameters and investigate numerically the economic growth and fluctuations in the model. We give a detailed account of the endogenous mechanism responsible for output fluctuations. We further study this mechanism in the Appendix, where we examine the structure of phase trajectories and identify the mechanism as coherence resonance. Section 4 ends the paper with a summary of conclusions.

## 1. Dynamic market model

First, we provide a simplified derivation of the model equations. In the process, we obtain a useful result, namely we specify the conditions that uniquely determine the parameters of microscale interaction – leading to the macro-level equations (Section 1.1). Second, we submit this model to empirical validation (Section 1.2). Finally, we demonstrate that the model contains the efficient market as a particular case (Section 1.3).

### 1.1. Derivation of model equations

Gusev et al. (2015) introduced an agent-based model in which investors' decisions are influenced by opinions exchanged with other investors and by opinions accessed through mass media. The latter are assumed to come from financial analysts, market commentators, newspaper journalists, finance bloggers and other participants who interpret relevant information, opine on how the market might react and make their views available to investors through media outlets (for convenience we refer to them collectively as analysts). That is, the model consists of two types of agents interacting at a micro level: investors whose role it is to trade and analysts who are responsible for information analysis.

Appendix A in Gusev et al. (2015) provides a formal treatment of this problem and detailed derivation of equations governing the evolution of macro variables. This section offers a simplified derivation to illustrate the modeling approach and enable the reader to develop intuition about the micro-macro connection in this model.

We begin by deriving a dynamic equation for investor sentiment (i.e. return expectation). Let us consider a large group of identical investors who can form differing binary opinions on the market direction. Let us also say that the $i$-th investor has the sentiment $s_i = +1$ if she opines that the market will rise and $s_i = -1$ if she opines that the market will fall. Then, the average sentiment $s$ per investor at time $t$ is given by [6]

$$s(t) = n_+(t) - n_-(t), \tag{1}$$

such that

$$n_+(t) + n_-(t) = 1, \tag{2}$$

---

[6] To clarify, the model contains a large number of investors, $N = (N_+ + N_-) \gg 1$, where $N_+$ and $N_-$ are the numbers of optimists and pessimists, respectively. Then, the proportions of optimists and pessimists are given by $n_+ = N_+/N$ and $n_- = N_-/N$. Note that $n_+(t)$ and $n_-(t)$ can be interpreted as the probabilities for an investor to have at time $t$ respectively a positive or negative market outlook.

where $n_+$ and $n_-$ are the proportions of the investors in the group who are, respectively, optimistic and pessimistic about the future market performance. By construction, sentiment $s$ can take any value between $-1$ and $1$.

Let us discuss the main forces that may influence sentiments of individual investors. First, by exchanging opinions, each investor aims to align the opposing investors along her directional view. At the leading order, we can treat this exchange as though sentiment $s_i$ is influenced by the average sentiment $s$ that acts to align $s_i$ along its direction.[7] Second, investors are also influenced by the analysts' binary market views disseminated through mass media. We can, as we have done above, take this influence to be determined in the leading order by the analysts' average expectation $h$ (we will discuss how to obtain $h$ later), noting that $h$ is defined per analyst and so varies between $-1$ and $1$. Hence, we express the total force $F$ acting on investors via interaction and dissemination as

$$F(s,h) = \beta_1 s(t) + \beta_2 h(t), \qquad (3)$$

where the positive constants $\beta_1$ and $\beta_2$ determine the corresponding sensitivities. Equation (3) implies that the more optimistic (pessimistic) the average expectations of investors and analysts are, the stronger the force that compels the pessimistic (optimistic) investors to reverse their views.

Next, we must account for a multitude of idiosyncratic factors that may affect investors. We assume that these factors act as random disturbances causing investors to occasionally change their views irrespective of other investors. In other words, the expectations of individual investors are assumed to be subject to random fluctuations.

---

[7] This treatment, called the mean-field approximation, is the leading-order approximation for a general interaction topology and so is a sensible first step for approaching the problem.

Consequently, the model requires a statistical description. Let us introduce $p^{-+}$ as the probability per unit of time for any pessimistic investor to switch her sentiment from $-1$ to $1$ and $p^{+-}$ as the probability per unit of time of the opposite change, which we henceforth call transition rates. The transition rates depend on the force $F$: they are equal when $F$ is zero and skewed by any nonzero $F$ such that $p^{-+} < p^{+-}$ for $F < 0$ and $p^{-+} > p^{+-}$ for $F > 0$.

Having defined $p^{-+}$ and $p^{+-}$, we are able to write down equations for the changes in $n_+$ and $n_-$ over a time interval $\Delta t$:

$$n_+(t + \Delta t) = n_+(t) + \Delta t(n_-(t)p^{-+}(t) - n_+(t)p^{+-}(t)), \tag{4a}$$

$$n_-(t + \Delta t) = n_-(t) + \Delta t(n_+(t)p^{+-}(t) - n_-(t)p^{-+}(t)). \tag{4b}$$

We proceed by rewriting equations (4) in terms of the average sentiment $s$. We first use equations (1) and (2) to obtain $n_+ = (1 + s)/2$ and $n_- = (1 - s)/2$ and then subtract equation (4b) from (4a) to arrive in the limit $\Delta t \to 0$ at the following differential equation for $s$:

$$\dot{s} = (1 - s)p^{-+} - (1 + s)p^{+-}, \tag{5}$$

where the dot denotes the derivative with respect to time.

To complete the derivation, we must find the transition rates $p^{-+}$ and $p^{+-}$ which are, as discussed above, some functions of the force $F$. We can do this by considering equations (4) in the state of equilibrium, $n_\pm(t + \Delta t) = n_\pm(t)$, which entails

$$\frac{p^{-+}}{p^{+-}} = \frac{n_+^0}{n_-^0}, \tag{6}$$

where $n_+^0$ and $n_-^0$ correspond to $n_+$ and $n_-$ at equilibrium.

Equation (6) is useful because it will interconnect the transition rates if we ascertain how the ratio of the optimist- and pessimist proportions in equilibrium, $\rho(F) = n_+^0/n_-^0$, depends on $F$. It may be simplest to assume that the relative change $d\rho/\rho$ is proportional to $dF$, i.e. $d\rho/\rho = \alpha dF$ with $\alpha$ being a positive constant. This assumption yields the equation:[8]

$$\frac{n_+^0}{n_-^0} = e^{\alpha F}. \tag{7}$$

It follows that $n_+^0 < n_-^0$ when $F < 0$, $n_+^0 = n_-^0$ when $F = 0$ and $n_+^0 > n_-^0$ when $F > 0$. Thus, as expected, the force acting on investors compels them to adopt a market outlook co-aligned with its direction, increasing the proportion of pessimists when it is negative or the proportion of optimists when it is positive as a result of the asymmetric transition rates.

Equations (6) and (7) entail the following condition: [9]

---

[8] This result can be obtained without making the above assumption on the basis that $n_+^0$ and $n_-^0$ must follow the Gibbs distribution (Gusev et al., 2015, Appendix A).

[9] By making use of equation (6) to derive condition (8), we implicitly assume that the same transition rates apply in- and outside of equilibrium, which is a standard approach in statistical mechanics. This assumption ensures that the equation determining the stationary values of sentiment *s* coincides with the equation for *s* in

$$\frac{p^{-+}}{p^{+-}} = e^{\alpha F}. \tag{8}$$

One additional constraint is required to determine $p^{-+}$ and $p^{+-}$ uniquely. To obtain it, we introduce the characteristic time $\tau_s$ over which random disturbances make individual sentiment $s_i$ flip. Recall that the transition rates are per unit of time, therefore $\tau_s$ represents the characteristic time over which the total probability that an investor's opinion reverses direction is equal to unity:[10]

$$(p^{-+} + p^{+-})\tau_s = 1. \tag{9}$$

Equations (8) and (9) yield the following expressions for the transition rates:

---

the state of thermodynamic equilibrium in the generalized Ising model in the statistical mechanics analog of this problem (Gusev et al., 2015, Appendix A).

[10] A sketch of its derivation is as follows. Consider a single interaction between two agents. As an agent has only two states, -1 and +1, her state after interaction will remain the same or change to the opposite. Introduce $p'^{++}$ and $p'^{+-}$ as the probabilities for the agent, initially in state +1, to be after interaction in state +1 or -1, respectively, and $p'^{--}$ and $p'^{-+}$ for the agent, initially in state -1, to be after interaction in state -1 or +1, respectively. We write $p'^{++} + p'^{+-} = 1$ and $p'^{--} + p'^{-+} = 1$ and assume that the agent's states after interaction do not depend on her initial state, i.e. $p'^{++} = p'^{-+}$ and $p'^{--} = p'^{+-}$. Thus we obtain $p'^{+-} + p'^{-+} = 1$. If interactions occur frequently on the timescale of interest, it is convenient to replace the discrete description by continuous. Then discrete probabilities per interaction $p'^{+-}$ and $p'^{-+}$ transform into probabilities per unit of time, i.e. transition rates $p^{+-}$ and $p^{-+}$, such as $p^{+-} = p'^{+-}/\tau_s$ and $p^{-+} = p'^{-+}/\tau_s$, where $\tau_s$ is the average time between interactions. And this yields equation (9).

$$p^{-+} = \frac{1}{\tau_s(1 + e^{-\alpha F})}, \tag{10a}$$

$$p^{+-} = \frac{1}{\tau_s(1 + e^{\alpha F})}. \tag{10b}$$

Thus, $p^{-+}$ and $p^{+-}$ are positive, bounded and, respectively, monotonically increasing and decreasing functions of $F$. As expected: $p^{-+} < p^{+-}$ for $F < 0$, $p^{-+} = p^{+-}$ for $F = 0$ and $p^{-+} > p^{+-}$ for $F > 0$.

We now use equations (10) and (3) to rewrite equation (5) that governs the evolution of the average investor sentiment $s$ as follows: [11]

$$\tau_s \dot{s} = -s + \tanh(\beta_1 s + \beta_2 h), \tag{11}$$

---

[11] Gusev et al. (2015) carried out a rigorous derivation of the evolution equation (11) in the context of the generalized Ising model proposed by these authors. The derivation presented here for illustration purposes incorporates certain assumptions made subject to the requirements of being both realistic and consistent with the formal derivation. In particular, we have obtained conditions (8) and (9) that determine the asymptotically bounded transition rates (10). Transition rates in this form appeared, for example, in Suzuki and Kubo (1968) – the first to derive the evolution equation (11) for a statistical physics problem. However, following Weidlich and Haag (1983), the interactions-based models in socioeconomic problems have mostly utilized transition rates that do not satisfy condition (9). Such transition rates, which yield the evolution equation different from equation (11), may be unrealistic as they are asymptotically unbounded. In addition, the transition rates given by equations (10) patch up the differences, discussed in Franke and Westerhoff (2017), between the transition probability (Weidlich and Haag 1983, Lux 1995) and discrete choice (Brock and Hommes 1997, 1998) approaches to modeling opinion dynamics.

where $\alpha/2$ has been absorbed into $\beta_1$ and $\beta_2$ without loss of generality. We remind that $\tau_s$ has the meaning of the characteristic time over which a sentiment $s_i$ flips and thus indicates the investor's average memory timespan. At the macro level, $\tau_s$ can be interpreted as the characteristic time of the variation in the investors' average sentiment.

The next step is to find how the analysts' average expectation $h$ evolves with time. The above derivation for $s$ can be applied here to lead to the dynamic equation for $h$ in the same form as (11) but with the force specific to $h$ in the argument of the hyperbolic tangent. Let us define this force. In our context, analysts translate news into opinions that influence investors, providing a channel through which exogenous information enters the model. Earlier we have discussed the special role of information related to price changes in the news-price-news feedback dynamic. Hence, we should separate such information from the rest of the general news flow. Next, it is also reasonable to assume that direct interactions are less important for analysts than for investors because time to market, from a news event to publication, is relatively short. Thus, for simplicity, we wish to neglect at the leading order the impact of interactions on opinion making in comparison with the impact due to exogenous information.

Based on these assumptions, we obtain the following equation for $h$:[12]

---

[12] Gusev et al. (2015) derived equations (11) and (12) together as a single system from the micro-level investor-analyst model (the generalized Ising model). Note that equation (11), with $h$ as an exogenous variable, was originally obtained by Suzuki and Kubo (1968) using the mean-field approach in a statistical mechanics problem. Subsequently, equation (11) has appeared in the socioeconomic context, often in stationary form. See, for example, reviews by Brock and Durlauf (2001) and by Franke and Westerhoff (2017).

$$\tau_h \dot{h} = -h + \tanh(k_1 \dot{p} + \zeta(t)), \tag{12}$$

where $p$ is the log price, $k_1$ is a positive constant that determines the feedback strength and $\tau_h$ represents the analysts' average memory timespan at the micro level and the characteristic response time at the macro level. The function $\zeta(t)$ denotes the flow of exogenous news comprising central bank announcements, corporate news, economic data release, political events and other relevant information. Gusev et al. (2015) treated $\zeta(t)$ as a stochastic exogenous variable to study stock market dynamics. In Section 2, we will make further assumptions on $\zeta(t)$ relevant for the study of economy-market interaction.

To close the model, we need a relation between expectation $s$ and price $p$. Investment expectations are usually equated with investment decisions, i.e. price changes are taken to be proportional to the difference between the number of optimists and pessimists in a model. However, this widely applied approach is not necessarily correct. Gusev et al (2015) alternatively suggested that investors act on their opinions differently over the short and long term, as defined relative to the average memory horizon $\tau_s$. Their basic argument is reproduced in the next paragraph.

Let us consider an investor who has just allocated capital to the market in line with her expectations. The following day, all else being equal this investor will not amend her allocation unless her sentiment changes because she has already deployed capital reflecting that same level of sentiment. Therefore, ignoring external constraints, the investment process over time horizons where the investors' memories of past sentiment levels persist ($t \ll \tau_s$) must be driven primarily by the change in sentiment. Conversely, over longer horizons ($t \gg \tau_s$) investors would invest or divest mainly depending on the level of sentiment itself because their previous allocation decisions would not be linked in their memory to previous levels of sentiment. Since capital flows cause price changes, these two asymptotic views, $\dot{p} \sim \dot{s}$ for $t \ll \tau_s$ and $\dot{p} \sim s$ for $t \gg \tau_s$, can be superposed (for the lack of simpler alternatives) to yield an equation that approximately describes the relation between price and expectation:

$$\dot{p} = c_1 \dot{s} + c_2 (s - s_*), \qquad (13)$$

where the constants $c_1$ and $c_2$ are positive and the constant $s_*$ can take any sign.

The phenomenological relation (13) states that price changes proportionally, first, to the change in investor sentiment and, second, to sentiment deviation from a certain reference level $s_*$ which can be interpreted as investors' expectations about the long-run economic growth fueled by technological progress. As discussed above, the first term is the main source of short-term price variation, while the second term determines leading behavior over long-term horizons.

Equations (11-13) form a nonlinear dynamical system that describes the market path as a trajectory evolving in the phase space $(h, s, p)$. Gusev et al. (2015) investigated the geometry of trajectories in this model to highlight the mechanics of certain market behaviors such as the transition between bull and bear markets. That paper also replicated, using empirically measured $h$, the actual price path and return distribution of the US stock market with relatively good precision. Kroujiline et al. (2016) derived a more general version of equations (11-13) to encompass investor groups with different investment horizons, thereby improving the model's precision, and applied this extended version for return prediction.

## 1.2. Empirical verification

This section augments the empirical studies in Gusev et al. (2015) and Kroujiline et al. (2016): it briefly reports the updated results of the empirical replication of market price path.

The empirical application is based on the measurement of $h$. Since $h$ is defined as the overall publicly expressed expectation regarding the direction of future market performance, this variable can be measured directly by parsing publicly available information. To this end, we seek $h$ as the ratio of the number of news items containing positive return expectations minus the number of news items containing negative return expectations over the total number of relevant news items.

The US stock market is selected as the object of this empirical study and the S&P 500 Index is taken as its proxy. We consider only the English language media. We apply the rule-based parsing methodology set out in Gusev et al. (2015) to daily news data for the period 1995-2017 retrieved from the DJ/Factiva news archive that comprises newspaper and journal articles, newswires, blogs and other publications.

As a result, we obtain a time series of daily $h(t)$ over this period (Figure 1a).[13] We substitute $h(t)$ into equation (11) and solve it numerically to obtain empirical investor sentiment $s(t)$ in Figure 1b. Then, we calculate model price $p(t)$ from $s(t)$ using equation (13). The resulting empirical model price $p(t)$ is shown in Figure 1c.

---

[13] More details about the measurement of $h$ are found in Gusev et al. (2015). In practice, as discussed there, news about current and recent market returns, which make up the bulk of the study's relevant news volume, are also included in the measurement of $h$, along with information about anticipated returns. This empirical approach is consistent with the price feedback concept implemented in the model via equation (12) as well as with empirical evidence that investors are likely to form expectations by extrapolating past returns (Greenwood and Shleifer, 2014).

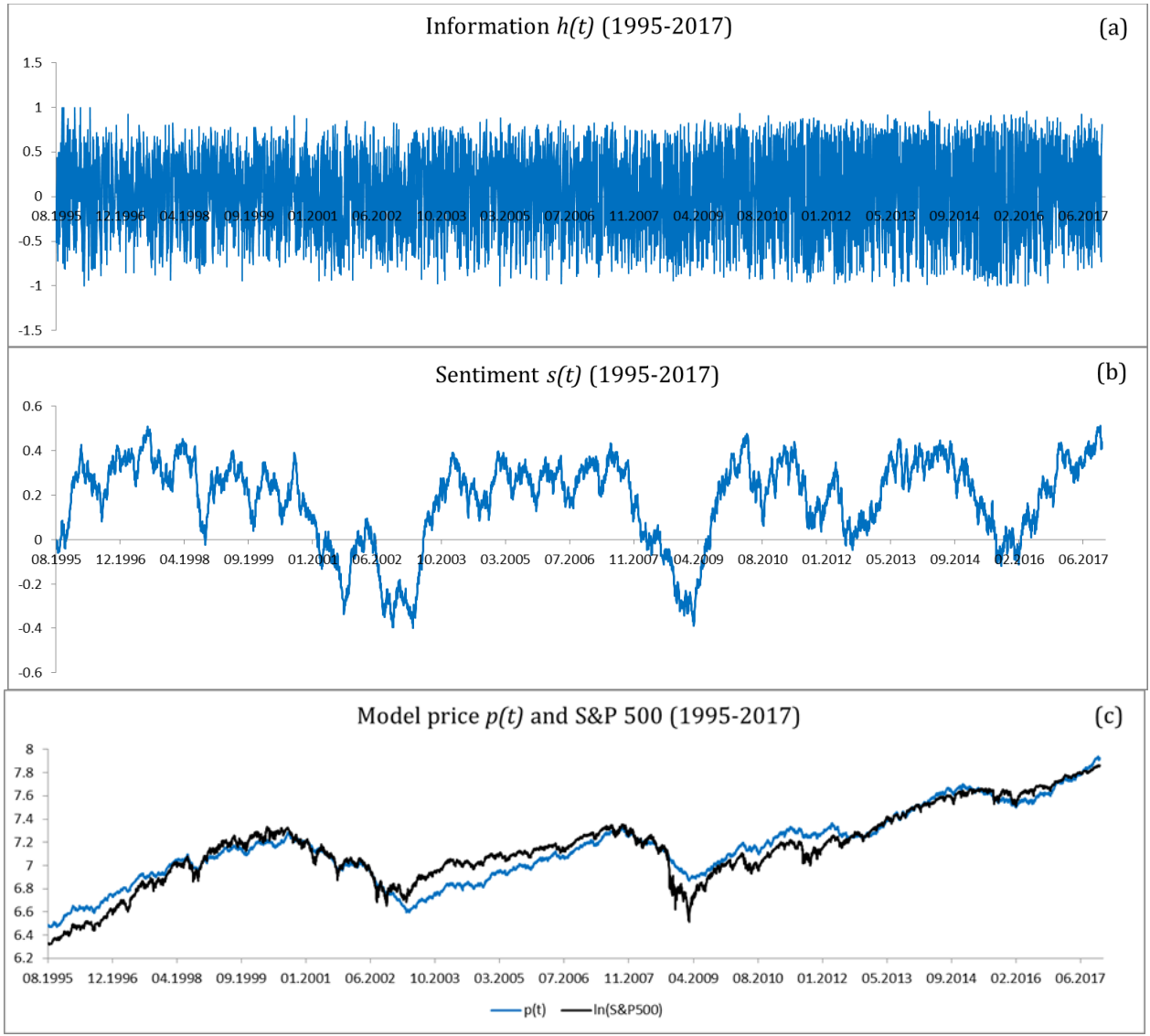

**Figure 1**: Daily time series of the empirical $h(t)$, $s(t)$ and $p(t)$ from 1995 to 2017. (a) Information $h(t)$ is measured using rule-based text parsing. (b) Sentiment $s(t)$ is calculated from measured $h(t)$ using equation (11) with the coefficients $\beta_1 = 1.1$, $\beta_2 = 1.0$ and $\tau_s = 25$ (business days) determined from realistic estimates and/or dominant balance considerations (see Section 3.1 for further explanations). Note that Gusev et al. (2015) demonstrated by an iterative minimization process that the coefficient values, most notably $\beta_1$, remain stable over time. (c) Model price $p(t)$ is calculated from $s(t)$ using equation (13) with the coefficients estimated by least squares fitting as follows: $c_1 = 0.337$, $c_2 = 0.003$, $s_* = 0.113$ and the integration constant equal to 6.421.

The correlation between the daily model prices and the daily index log prices is 95%. Thus, the model can credibly replicate the past price track.

It is outside the scope of this paper to test the model's predictive power. We only note that Kroujiline et al. (2016) developed, using the generalized version of equations (11-13), a return forecast methodology based on the analysis of the market position in the model's phase space and constructed an investment strategy upon it. This strategy, which has news as its only input, has been live traded since April 1, 2016.

## 1.3. Efficient market limit

According to model (11-13), the stock market evolution is driven by the exogenous news flow and by the endogenous price feedback mechanism. The former contains a stochastic component and therefore causes the market price to move randomly. The latter induces internal dynamics, giving rise to deterministic behaviors.

It would be instructive to consider the situation where the market path is determined primarily by stochastic influences. In this case, the market will exhibit a random walk as it would if it were efficient. Therefore, we can expect that model (11-13) includes the efficient market regime as a particular case. Let us discuss how this case can be recovered.

To capture market efficiency, we will make the following assumptions in model (11-13): no interaction among investors ($\beta_1 = 0$); no price feedback ($k_1 = 0$); analysts interpret information instantaneously as it arrives ($\tau_h = 0$); and investors react immediately on the opinions supplied by analysts ($\tau_s = 0$). We treat $\zeta(t)$ as a stochastic news flow and also assume that, over the long term, the volume of positive news is on average greater than the volume of negative news as a reflection of technological progress. Therefore, we set $\zeta(t) = \varepsilon + \xi_t$ where $\varepsilon > 0$ is the constant mean and $\xi_t$ is a normally-distributed zero-mean white noise. We further assume that the news flow fluctuations as well as the mean are small, $|\xi_t| \ll 1$ and $\varepsilon \ll 1$, thereby excluding the extreme market regimes such as crashes and rallies accompanied by high news activity, where the market is most likely behaving inefficiently.

These assumptions reduce equations (11) and (12) in the first order respectively to $s = \beta_2 h$ and $h = \zeta(t) = \varepsilon + \xi_t$, while the assumption $\tau_s = 0$ implies that the equation of price formation (13) must be amended to include only the asymptotic $c_2(s - s_*)$ valid at $t \gg \tau_s$ (Section 1.1). Thus, the model equations (11-13) become

$$s = \beta_2 h, \qquad h = \varepsilon + \xi_t, \qquad \dot{p} = c_2(s - s_*), \tag{14}$$

which after integration leads to the random walk of market price:

$$p(t) = c_2 \beta_2 \int \left(\xi_t + \left(\varepsilon - \frac{s_*}{\beta_2}\right)\right) \mathrm{d}t. \tag{15}$$

Note that the price drift is positive provided $\varepsilon > s_*/\beta_2$, which means that, in the long run, good news reflective of technological progress must (approximately) exceed expectations to sustain the long-term growth.

Hence, as expected, model (11-13) does not refute the efficient market model but instead encompasses it as a particular case. To sum up, the results reported so far enable us to reasonably conclude that model (11-13), derived from first principles, establishes a general, tractable and credible framework for explaining market behaviors. In the next section, we will apply this model in a macro-economic context to highlight endogenous mechanisms that may be responsible for producing business cycles and facilitating economic growth.

## 2. Basic real economy – stock market model

Our objective is to investigate the key effects that the stock market model (11-13) can generate through interaction within a macroeconomic system. For this purpose, we apply a basic relation between output and demand, stripped of any auxiliary features unnecessary for coupling the economy and the market. Such a relation was proposed by Blanchard (1981) – being an extension of the classic IS equation – which sets stock market price to be the main factor (besides income) driving spending via investment (in the sense of Tobin's q) and consumption (wealth effect):

$$\tau_y \dot{Y} = Y_D(Y, P) - Y = aP - bY, \tag{16}$$

Equation (16) states that output $Y$ adjusts over time $\tau_y$ to any changes in demand $Y_D$ which is proportional to income, i.e. output $Y$, and stock market price $P$.[14] For convenience, we express it in terms of the log output $y = \ln Y$ and the log price $p = \ln P$ (while absorbing $1/a$ into $\tau_y$ and $b$):[15]

$$\tau_y \dot{y} = e^{p-y} - b. \tag{17}$$

---

[14] Blanchard (1981) also included a fiscal policy term. We do not consider policy effects in this work.

[15] Equation (17) can be obtained, as a limiting case, from a dynamic extension of Cobb-Douglas production, where output adjusts over time to changes in capital subject to credit frictions (Gusev et al, 2019).

Let us consider equations (11-13) and (17) together. In this macroeconomic system, the stock market affects the real economy via equation (17). We must make assumptions on how the economy can in turn influence the market in order to close this system. Incidentally, the news flow $\zeta(t)$, being the only exogenous variable in the market model (11-13), provides such a connection. Recall that $\zeta(t)$ comprises any public information that may be relevant to forming an investment opinion. It is reasonable to expect that, in a growing economy, the volume of positive economic news (such as those concerning company earnings or economic indicators) would on average exceed the volume of negative news and, conversely, negative news would on average prevail in a contracting economy. Further, as discussed in Section 1.3, we expect that, in the long run, $\zeta(t)$ is positively shifted to reflect technological progress. As such, we can divide the news flow $\zeta(t)$ into three parts:

$$\zeta(t) = k_2 \dot{y}(t) + \varepsilon + \xi_t, \tag{18}$$

where $k_2 \dot{y}(t)$ is its variation in line with the rate of change of economic output; $\varepsilon$ is its long-term positive mean related to technological progress; and $\xi_t$ is the remainder that we assume to be normally-distributed white noise with zero mean.

As a result, we obtain a self-contained dynamical system for the coevolution of the general economy and the stock market:

$$\tau_y \dot{y} = e^{p-y} - b, \tag{19a}$$

$$\dot{p} = c_1 \dot{s} + c_2(s - s_*), \tag{19b}$$

$$\tau_s \dot{s} = -s + \tanh(\beta_1 s + \beta_2 h), \tag{19c}$$

$$\tau_h \dot{h} = -h + \tanh(k_1 \dot{p} + k_2 \dot{y} + \varepsilon + \xi_t). \tag{19d}$$

Equation (19a) defines a very basic economic relation that we apply to study the economy-market interaction. Nonetheless, the stock market model (19b-d), possessing nontrivial internal dynamics, sets in motion complex endogenous behaviors in the coupled system, as we will see, even where the attached economic component is as simple as (19a). It should be possible to use equation (19d) to pair the stock market model (19b-d) with other economic models, so that this stock market model could be utilized in various relevant economic settings.

In the present work, we prioritize simplicity in the choice of the economic component in the coupled system in order to focus on the quintessential effects arising through interaction. We will discuss two such effects captured and explained by equations (19): the short- to mid-term economic fluctuations and the long-term economic growth. These results require certain clarification. Since relation (19a) implies that economic prices remain fixed, it is usually interpreted as applicable on timescales much shorter than the timescale of price adjustment. Here, given the above-stated priority, we use this basic relation as an instructive and tractable example for illustration purposes. Therefore, we choose to interpret it as describing a hypothetical economy where inflation is absent irrespective of timeframe. As such, we will additionally study equations (19) over long time horizons.

This paper aims to contribute to the existing research on business cycles by exploring new ideas on the interaction between the real economy and the stock market. We therefore conclude this section by briefly outlining the context for the present work. Equation (19a) is a convenient starting point. It was proposed by Blanchard (1981) for integrating a rational-expectations stock market model into a coupled economy-market system. Subsequent research has been focused on extending Blanchard's work in order to enhance the economy's dynamics via its mutual interaction with the market. In particular, agent-based market models with heterogeneous investors (namely, the familiar noise- and fundamental traders) have been introduced into this framework to derive macroeco-nomic systems with multiple equilibria and endogenous dynamics, such as Chiarella et al. (2006) and, especially relevant in the present context, Franke and Ghonghadze (2014) and Flaschel et al. (2017) that involve investor opinion dynamics.[16]

The present work shares the motivation with the above research, as we also seek to explain business cycles as an endogenous phenomenon arising in the economy through interaction with the (dynamic) market. However, our approach differs from the existing literature in the following key aspects. First, the market model employed in this paper is substantially different from other models. It is microfounded in a different mechanism (Section 1.1) and is empirically validated (Section 1.2). Second, the feedback from the economy into the market is treated in a novel way by noticing that exogenous news flow must be modulated by the economy's performance (equation 18). Finally, the model's tractability enables the identification of the mechanism underlying business cycles and al-lows us to provide a detailed explanation of this mechanism using both real world (Section 3.3) and phase space (Appendix) descriptions.

---

[16] Further with respect to the relevant literature, we take note of Benhabib et al. (2016) that studies the role of sentiment-driven financial markets in business cycle formation in the context of the "sunspot" literature (see Farmer, 2014). It also views business cycles as endogenous fluctuations arising in a coupled economy-market system with multiple equilibria; however, their model and the model developed here differ greatly by construction and behavior. Also, it is important to mention the fast-growing body of research in agent-based computational economics, where out-of-equilibrium endogenous dynamics on the macro level are obtained via direct simulation of micro-level interactions among agents. See reviews by LeBaron and Tesfatsion (2008) and by Fagiolo and Roventini (2012) that includes a detailed comparison with the DSGE approach. Finally, as the news-driven market plays a central role in our study, it shares, generally speaking, common ground with the recent literature on the role of news in the formation of business cycles within the DSGE framework (e.g. Beaudry and Portier, 2004, 2014; Jaimovich and Rebelo, 2009).

## 3. Study of the coupled real economy – stock market model

This part of the paper is structured as follows. Section 3.1 establishes the realistic value ranges for the model's parameters. Section 3.2 examines the model's behavior in the long run. It highlights the connection between the stock market and the real economy over these time horizons and, in particular, the role that the market may play in translating news about technological advancement into the economic growth. Section 3.3 studies the mechanics of the business cycles generated by the model, explaining them as quasiperiodic fluctuations stemming from the endogenous interaction between the market and the economy. Additionally, the Appendix provides further insights into the mechanism behind these fluctuations by inspecting the model's phase space.

### 3.1. Realistic parameters

In this section, we review the model's parameters with a focus on finding their realistic values. Table I at the end of the section summarizes the values selected for studying the model.

First, there are the characteristic response times $\tau_h$, $\tau_s$ and $\tau_y$. The time $\tau_h$ cannot be longer than several days because analysts must transmit their views to the investment community without undue delay following a news event. We set $\tau_h = 2.5$ business days.[17] The characteristic times of change in investor sentiment are extremely heterogeneous. They vary within a broad range, from seconds to years depending on the investment horizons of relevant investor groups and strategies. We take $\tau_s = 25$ business days as an average estimate. The output adjustment time $\tau_y$ must correspond to the timescale of GDP variation. We assume that it is approximately four years, so that $\tau_y = 1000$ business days. It follows that the characteristic times in model (19) differ by roughly an order of magnitude: $\tau_h \ll \tau_s \ll \tau_y$.

---

[17] This value is consistent with the behavior of the serial correlation of empirical $h(t)$, which decays over the span of 1-3 business days.

Let us now establish the realistic value ranges for other parameters, starting with equation (19c). Parameter $\beta_1$ determines the relative importance of the herding and random behaviors of investors. The herding behavior prevails for $\beta_1 > 1$, whereas the random behavior is stronger for $\beta_1 < 1$. It is sensible to assume $\beta_1 \sim 1$, otherwise investors would unrealistically behave either in perfect synchronicity or randomly. We take $\beta_1 = 1.1$ as it provides a better fit with the empirical data (Gusev et al., 2015), implying a slight prevalence of herding over randomness.[18] Parameter $\beta_2$ determines the impact of analyst opinion on investors. As this impact is an important element of the model, $\beta_2 \sim 1$ is necessary for it to enter the leading-order balance.

In equation (19d), parameters $k_1$, $k_2$ and $\varepsilon$ control, respectively, the impacts of market price, economic output and technological progress on the generation of $h$. The short-term price feedback $k_1 \dot{p} \sim k_1 c_1 \dot{s}$ plays a central role in market dynamics because it can effectively couple $h$ and $s$.[19] Thus, since $|\dot{s}| \sim 1/\tau_s$ as follows from equation (19c), it is necessary that $k_1 \sim \tau_s$ to allow this term at the leading order, assuming $c_1 \sim 1$.

---

[18] Equation (19c) has a phase transition at $\beta_1 = 1$ that leads to a bifurcation of equilibrium points, such that provided $h = 0$ (analysts exert no force on investors) sentiment $s$ converges to a single stable equilibrium point at $s = 0$ for $\beta_1 < 1$ (random behavior) and to one of two stable equilibrium points at $s < 0$ or $s > 0$ for $\beta_1 > 1$ (herding behavior) depending on the initial conditions.

Unlike the market price, changes in economic output are not visible on a daily basis. The influence of the economic state is gradual: it works by incrementally affecting news sentiment over long periods of time. Therefore, we expect that $k_2 |\dot{y}| \ll 1$, which entails $k_2 \ll \tau_y$ since $|\dot{y}| \sim 1/\tau_y$ as follows from equation (19a). Similarly, we anticipate that technological progress does not appear in the leading-order balance in equation (19d) and therefore assume $\varepsilon \ll 1$.

We select the parameter values within the above-described ranges for which the modeled economic output approximately matches the long-run economic growth as well as exhibits realistic fluctuations with the amplitudes analogous to the observed business cycles. These values, reported in Table I, will be utilized in the numerical simulations in the next sections.[20]

---

[19] Since $h$ varies much faster than $s$ ($\tau_h \ll \tau_s$), the long-term feedback component $k_1 \dot{p} \sim k_1 c_2 (s - s_*)$ cannot make $h$ and $s$ "adhere" to each other strongly enough to generate robust, diverse dynamics at short timescales. So it is reasonable to expect the short-term feedback to dominate the long-term feedback in the leading-order balance: $k_1 c_1 |\dot{s}| \gg k_1 c_2 |s - s_*|$. This condition entails $c_2 \ll c_1/\tau_s$ as $|\dot{s}| \sim 1/\tau_s$ and $|s - s_*| \sim 1$, which is in line with the results of the empirical study in Section 1.2.

[20] The values of $c_1$, $c_2$ and $s_*$ are estimated using the US real GDP per capita.

**Table I**: Parameter values applied in the model (equations 19).

| Parameter | Value |
|---|---|
| $\tau_h$ | 2.5 (business days) |
| $\tau_s$ | 25 (business days) |
| $\tau_y$ | 1000 (business days) |
| $\beta_1$ | 1.1 |
| $\beta_2$ | 1.0 |
| $k_1$ | 30 |
| $k_2$ | 175 |
| $\varepsilon$ | 0.03 |
| $c_1$ | 1.0 |
| $c_2$ | 0.00022 |
| $s_*$ | 0.03 |
| $b$ | 0.5 |

## 3.2. Economic growth

Here we investigate the model's behavior over long time horizons, interpreting equation 19a as a hypothetical economy with fixed prices (see Section 2). To this end, we solve numerically equations (19), using the stochastic daily news flow component $\xi_t$ as an exogenous input variable[21], to obtain $h(t), s(t), p(t)$ and $y(t)$.

---

[21] On daily intervals, $\xi_t$ is modeled as normally-distributed white noise with zero mean and a standard deviation of 0.4. We have chosen $\xi_t$ to have a small positive intraday autocorrelation on the assumption that news events are positively correlated on intraday time intervals (the autocorrelation is zero over the intervals of one day and longer).

Figure 2a depicts a simulated path of economic output $y(t)$ over 222 years and the US real GDP per capita for the period 1790-2012. We can see that the modeled output experiences a long-term linear growth on a logarithmic scale and also exhibits a generally similar pattern and amplitude of fluctuations as the actual output.

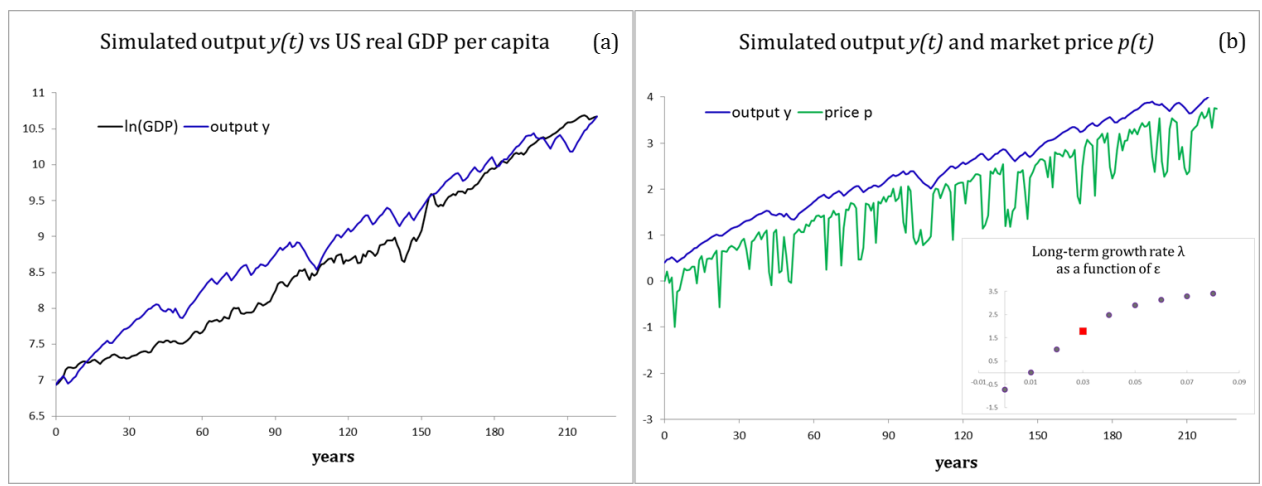

**Figure 2**: (a) A simulated path of output $y(t)$ vs. the US real GDP per capita 1790-2012 (logarithmic scale); the simulated output path is parallel shifted to match the value of the GDP at the start. GDP source: MeasuringWorth.com. (b) This same output path $y(t)$ and the price path $p(t)$ from the same simulation; the output path is parallel-shifted to stay above the price path for a better visualization. The inset panel shows the long-run output annual growth rate $\lambda$ (%) as a function of the news flow parameter $\varepsilon$ reflecting technological progress. Each point in the panel corresponds to $\lambda$ simulated for a given value of $\varepsilon$ and averaged over 1000 years; the red square denotes $\lambda$ for $\varepsilon = 0.03$ utilized in the numerical simulations in this paper (Table I).

Similarly, price $p(t)$ grows linearly at the same rate as $y(t)$ in the long run (Figure 2b). This must be expected: according to equation (19a), output adjusts to changes in price over time $\tau_y$ which is four years; as this time lag is negligible over the span of two centuries, output growth is closely tracking that of the market over the long term.

This growth is fueled by the news flow parameter $\varepsilon$ which reflects technological progress. The underlying mechanism is as follows. An $\varepsilon > 0$ causes $h$ to be on average positive (equation 19d) which in turn forces investor sentiment $s$ to also stay on average positive over long time horizons (equation 19c). This long-term sentiment connected to technological progress influences the long-term behavior of market price via the term $c_2(s - s_*)$ in equation (19b). In particular, the market will, in the long run, grow provided that this sentiment $s$ is greater than the expectation given by $s_*$. As a result, economic output will increase in tandem with market price (equation 19a) at the long-term growth rate $\lambda$. The inset panel in Figure 2b shows $\lambda$ as a function of $\varepsilon$.

We have conducted this study in a simplified economic framework given by equation (19a), where the market is the sole source of growth in economic output and which, as discussed above, cannot be realistically applied to long time horizons. This section, therefore, serves only the purpose of illustrating the effect of news as a channel through which technological progress may drive the economy; for example, similar effects were studied in Jaimovich and Rebelo (2009). In this context, according to model (19), the market plays the role of a transmission mechanism between the technological advancement reflected in news and long-run economic growth.

### 3.3. Business cycles

In this section we examine the output fluctuations around the straight line mean path – business cycles. For this purpose, we construct the histogram of the lengths of the business cycles simulated by the model and compare it to the histogram representing the real-world business cycles. We detrend the data using in both cases the same non-causal moving average and take the distance be-tween two successive troughs as a measure of the business cycle length. Figure 3 shows the resulting histograms.

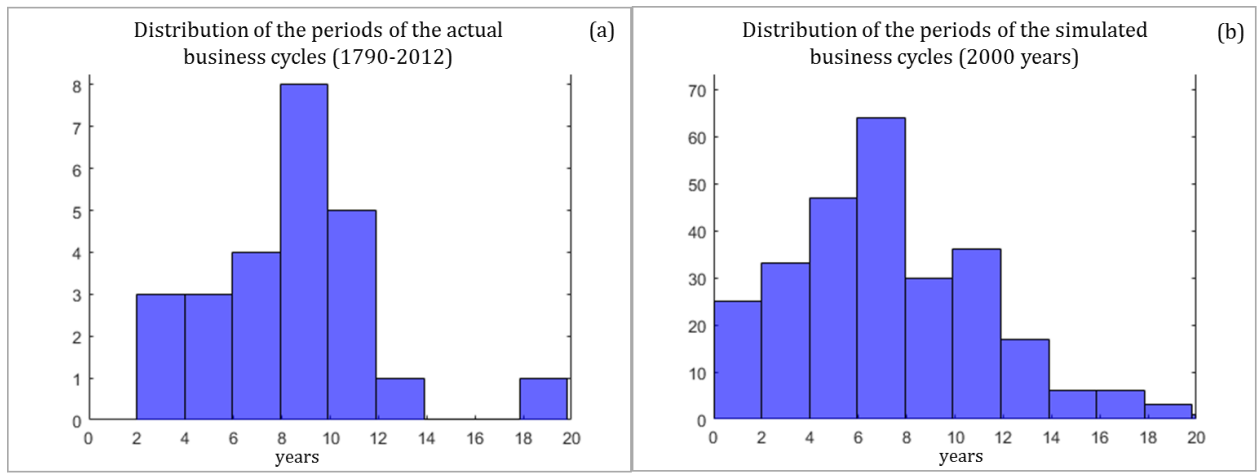

**Figure 3**: Histograms of the lengths of business cycles (a) observed and (b) simulated. The observed business cycles are obtained by detrending the US real GDP per capita 1790-2012 using the 12yr non-causal moving average and taking the distance between two successive troughs as a measure of length. The simulated cycles are obtained from a 2000yr output simulation with the same method used for observed cycles.

We can see that the distribution of the periods of the simulated business cycles displays a peak at 6-8 years, indicating the presence of quasiperiodic fluctuations. The distribution of the periods of the actual business cycles has a peak in the 8-10yr interval. Thus, we note that the model, with the chosen parameters, exhibits quasiperiodic fluctuations in output that correspond reasonably well with the observations.

The relative simplicity of the model enables us to identify the mechanism responsible for the generation of business cycles. This mechanism, called coherence resonance, occurs in stochastically-forced dynamical systems with variables evolving on different timescales – such as model (19). We will carry out this study in the Appendix, where we will discuss how this mechanism operates by examining trajectories in the model's phase space. Here, we would like to highlight its endogenous nature as a product of the interaction between the economy and the market.

Figure 4 provides an expanded view of two business cycles taken from the 222yr evolution path simulated by model (19) (Figure 2). Economic growth $\dot{y}$, market price $p$ and investor sentiment $s$ are plotted as functions of time with the time intervals of characteristic behaviors are indicated. The economy undergoes expansion ($\dot{y} > \lambda$) in intervals A-B and contraction ($\dot{y} < \lambda$) in intervals B'-A' relative to its long-term growth rate $\lambda$, while the transition from expansion to contraction and vice versa occurs respectively in intervals B-B' and A'-A. As in the real world, the simulated business cycles differ in amplitude and duration. In addition, the unique patterns of shorter-term fluctuations specific to each business cycle are clearly visible in the graphs.

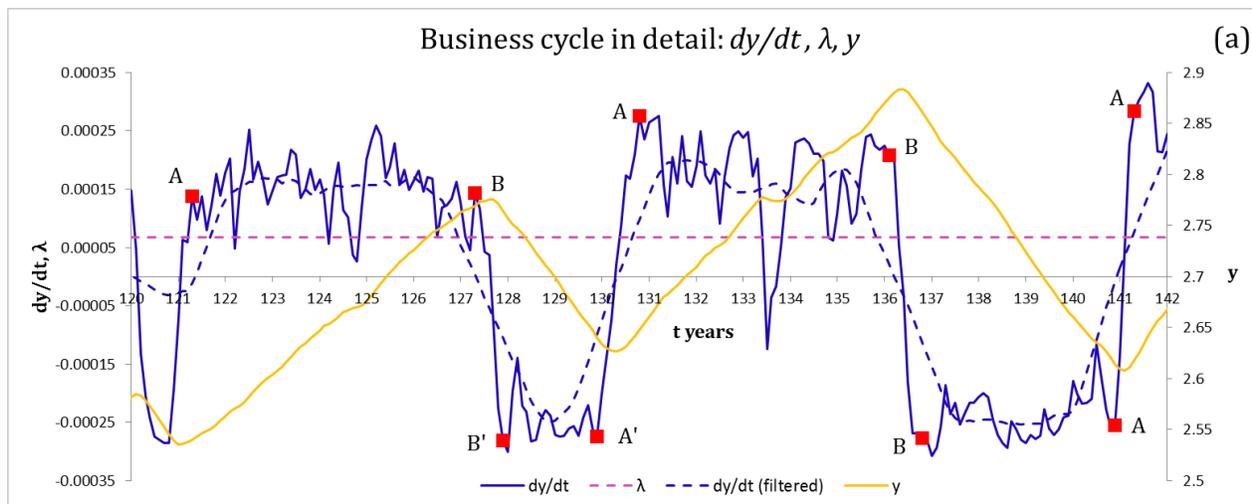

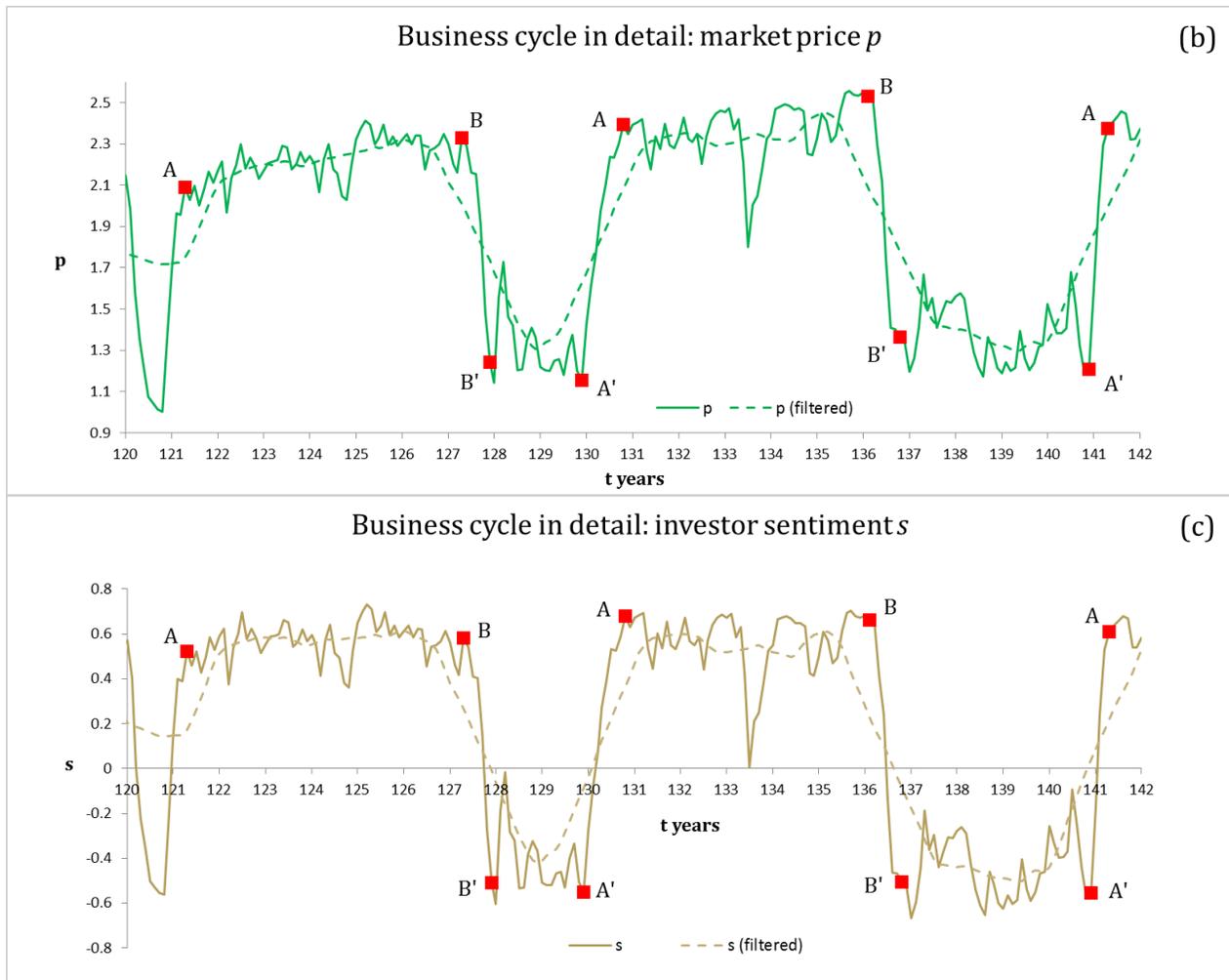

**Figure 4**: Two successive business cycles from the 222yr output simulation in Figure 2. The periods of characteristic regimes are separated by red squares. (a) Output growth $\dot{y}(t)$, actual and filtered using the 2yr non-causal moving average; output $y(t)$; and the long-run output growth rate $\lambda$. (b) Price $p(t)$, actual and filtered using the 2yr non-causal moving average. (c) Sentiment $s(t)$, actual and filtered using the 2yr non-causal moving average.

Let us turn our focus to the second business cycle in Figure 4. It starts in year 131 as sentiment and price, having rallied from the lows of the bear market, enter the bull market phase. The investors' outlook represented by sentiment stays steadily positive during this phase for five years.

This may seem strange as sentiment, being a bounded variable, could be expected to commence its reversal after reaching a peak, signaling an overvalued market.

What is stopping sentiment from deteriorating for so long? The answer lies in the behavior of output. Let us see how. The preceding market rally (year 130) instigates an enduring growth in output between years 131 and 136. This growth supplies a stream of positive news ($k_2 \dot{y}$ in equation 19d) that presents a barrier for a random adverse news event ($\xi_t$ in equation 19d) to make a material impact on sentiment. Thus, investor sentiment becomes effectively captive in the vicinity of its peak value.

This positive sentiment creates an extended upward price trend, i.e. a bull market. In this regime, however, price grows at a slower rate than during the market rally. This is because the bull market and the market rally have different causes. According to equation (19b), the rally is driven by $c_1 \dot{s}$ since the change in sentiment is the dominant dynamic in this fast-paced regime, while the second-order magnitude $c_2(s - s_*)$ is responsible for the bull market trend because the sentiment variation in this phase is restricted by the entrenchment mechanism described above. Equation (19a) ensures that the growth in output adjusts to this change in pace; consequently, the rate of economic expansion $\dot{y}$ gradually decreases.

As $\dot{y}$ continues to decrease, the barrier that a random negative news event must overcome to shift the entrenched sentiment down becomes smaller. As a consequence, the probability of such a news event gradually increases. This event, which occurs in year 136, appears as a stochastic exogenous shock. It triggers a negative spike in $\dot{h}$ that propagates through $\dot{s}$ on to $\dot{p}$. The latter activates price feedback ($k_1 \dot{p}$ in equation 19d) that acts to amplify the downward movement, causing sentiment and price to plunge together into the bear market.

Once in the bear market, sentiment again becomes captive, this time with the opposite sign. The economic contraction ensues following the market crash. It feeds the negative news flow that keeps investor sentiment in a trough between years 137 and 141. As the contraction subsides, the barrier that prevented an occasional positive news event from lifting sentiment diminishes with increasing $\hat{y}$. Eventually, a positive random news event sets in motion the market rally in year 141. The business cycle has thus come full circle.

Summing up, model (19) attributes the emergence of business cycles to the interaction between the real economy and the stock market. In this interaction, the economy imposes its slow timescale on the coupled system by, figuratively speaking, closing and opening the window of opportunity for a regime change. The probability of change is higher in the later stages of economic contraction and expansion as market sensitivity to regime-contradicting news gradually increases. Eventually, a random news event succeeds in activating the news-price-news feedback dynamic that drives transitions between the bear- and bull markets. The ensuing market rally (crash) propels the economy from contraction (expansion) to expansion (contraction). It follows that the economy and the market influence each other on different timescales creating nontrivial endogenous dynamics that involve deterministic and stochastic features. As a result, business cycles emerge as quasiperiodic fluctuations in output of varying durations, the distribution of which has a distinct maximum (Figure 3). This endogenous mechanism is identified as a coherence resonance in the Appendix.

## 4. Conclusion

This paper conjectures that business cycles may be a manifestation of the economy's endogenous dynamics, develops the relevant dynamic model and explains an endogenous mechanism underlying economic fluctuations consistent with the observed business cycles. The motivation behind this study is as follows. Certain behaviors of the stock market, such as crashes and rallies, point to the presence of feedback mechanisms – a plausible source of stock market dynamics. These dynamics may induce, through interaction, adjustments in the slow-moving real economy and, in turn, be influenced by these adjustments. As a result, the real economy – stock market interaction can cause endogenous dynamics in the coupled system across multiple timescales, including the fluctuations at business cycle frequencies.

The following contributions have been made in the course of this study:

1. We presented a simplified microfounded derivation of the agent-based stock market model of Gusev et al. (2015). This derivation sets out conditions that uniquely determine the transition rates in microscale interactions, leading to the macro-level equations that describe the stock market dynamics in terms of the interaction between information flow, investor expectation and market price subject to the impact by exogenous news.
2. We provided further empirical evidence in support of the model's validity and demonstrated that the model contains the efficient market regime as a special case.
3. We proposed that this market model can be embedded into the macroeconomic framework via exogenous news, incoming into the market, because the news flow is modulated by the economic growth. To illustrate key effects, we linked it with a simple dynamic extension of the IS equation where stock market price directly impacts aggregate demand (Blanchard, 1981). As a result, the real economy and the stock market become mutually coupled via news and price in this system.
4. The study of this coupled economy-market system showed that the economy and the market influence each other on different timescales, creating nontrivial endogenous dynamics that involve deterministic and stochastic features. In particular, the economy-market interaction generates quasiperiodic fluctuations that exhibit a peak in their length distribution consistent with that of the actual business cycles. This system also highlights the potential role of the market as a transmission mechanism between technological progress, reflected in news, and long-run economic growth.

5. This study further shows that the coupled economy-market system has two stable equilibria in the phase space. One stable equilibrium corresponds to the bull market and the expanding economy and the other to the bear market and the contracting economy. The system evolves dynamically under the action of the endogenous forces pulling it toward one of the two equilibrium states and the stochastic force exerted by the flow of exogenous news that excites it and triggers regime changes. The influence of technological progress, reflected in exogenous news, compels the system to stay on average longer in the vicinity of the economy expansion and bull market equilibrium, resulting in the long-run economic- and market growth. Finally, the mechanism of economic fluctuations was identified as coherence resonance – a phenomenon occurring in stochastically-forced dynamical systems (Pikovsky and Kurths, 1997).

We finish by noting that the incorporation of stock market dynamics into the macroeconomic framework, as presented here, allowed us to obtain realistic economic- and market behaviors within a tractable setting, so that this approach can potentially enhance models applied for policy analysis.

**Acknowledgments**

We are grateful to LGT Capital Partners for partially funding this project. We would also like to thank John Orthwein for editing this paper and contributing ideas on its readability.

**Appendix: Business cycles as coherence resonance**

Here, we explain the mechanism underlying the quasiperiodic fluctuations in terms of the structure of phase trajectories.

As the first step, we introduce a new variable $z$ such that

$$z = p - y. \tag{20}$$

We expect $z$ to be bounded because in the long run $p$ and $y$ grow with time asymptotically at the same rate (Section 3.2). This change of variables transforms model (19), which admits the infinitely growing solutions, into a dynamical system with bounded phase trajectories:[22]

$$\dot{z} = c_1 \dot{s} + c_2(s - s_*) - \omega_y(e^z - b), \tag{21a}$$

$$\tau_s \dot{s} = -s + \tanh(\beta_1 s + \beta_2 h), \tag{21b}$$

$$\tau_h \dot{h} = -h + \tanh\bigl(k_1(c_1 \dot{s} + c_2(s - s_*)) + k_2 \omega_y(e^z - b) + \varepsilon + \xi_t\bigr), \tag{21c}$$

where $\omega_y = 1/\tau_y$ for convenience.

System (21) possesses three equilibria for the parameters in Table I (Figure 5). Two equilibrium points are stable focus-nodes: one equilibrium located in the region where $s < 0$ and the other where $s > 0$. The former corresponds to the bear market and the contracting economy, while in the latter the market and the economy are growing. The third equilibrium point is an unstable saddle located between the two stable equilibria. In the absence of external news ($\xi_t = 0$), the economy-market system would converge to the state of stable equilibrium where, depending on the initial conditions, the economy is either contracting ($s < 0$) or expanding ($s > 0$). The nonzero $\xi_t$ does not let the system settle at these equilibria; therefore the market and the economy continue to coevolve dynamically in accordance with equations (21).

---

[22] Note also that the system's motion is bounded in $s$ because $-1 \leq s \leq 1$ while $\dot{s} > 0$ at $s = -1$ and $\dot{s} < 0$ at $s = 1$, as follows from equation (19c) or (21b), and is similarly bounded in $h$ because $-1 \leq h \leq 1$ while $\dot{h} > 0$ at $h = -1$ and $\dot{h} < 0$ at $h = 1$, as follows from equation (19d) or (21c).

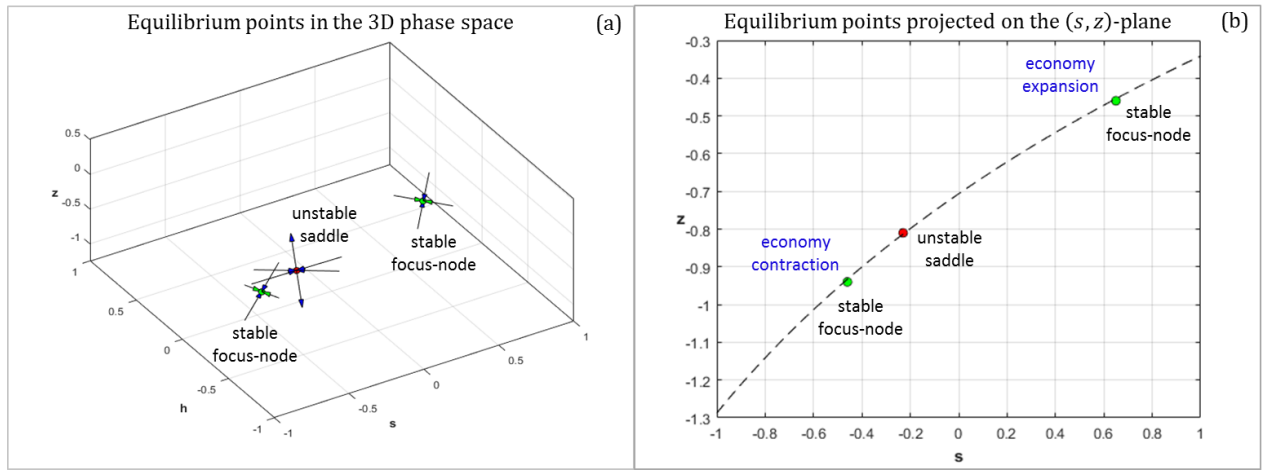

**Figure 5**: The equilibrium points of dynamical system (21): (a) in the phase space $(h, s, z)$ and (b) as projected on the $(s, z)$-plane. The equilibrium points comprise two stable focus-nodes, where one point corresponds to economy contraction and the other to expansion, and one unstable 3D saddle between the stable equilibria. Note that the projections of the equilibrium points are located on the line $z = \ln\left(b + \tau_y c_2 (s - s_*)\right)$.

The coevolving economy-market system is akin to a particle tracing a path in the phase space $(h, s, z)$. Figure 6a depicts a 14-year evolution path simulated by equations (21), which covers two business cycles. We can observe that the system spends most of its time in the neighborhood of the stable equilibrium points, which act as attractors, and crosses from one attracting region to the other infrequently.

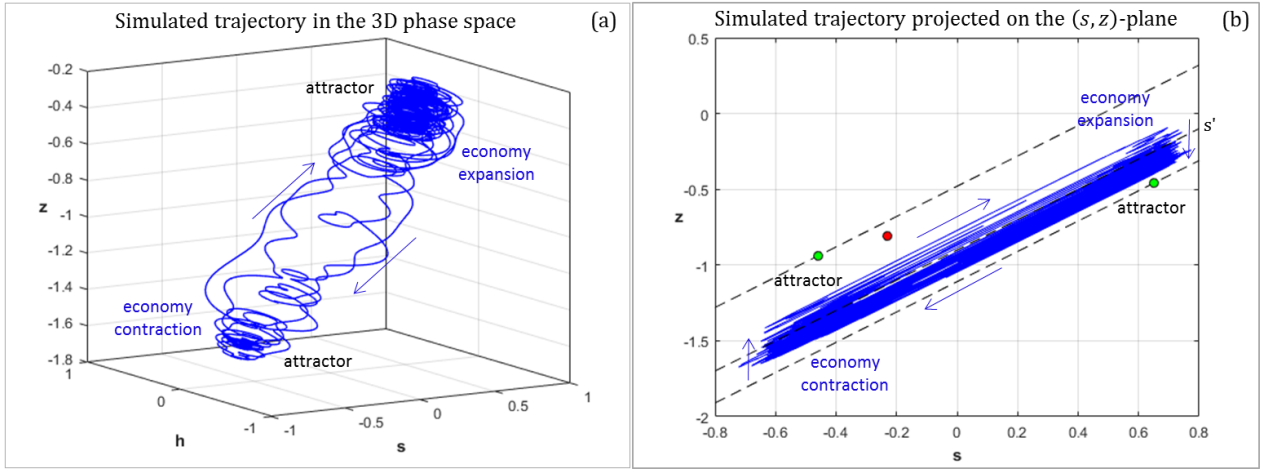

**Figure 6**: Simulated trajectories. (a) A 14yr evolution path smoothed by a Fourier filter to remove harmonics with periods less than 10 business days for a better visualization in the 3D phase space. Two attracting regions formed by the stable equilibrium points are visible. (b) The 222yr path from Section 3.2 projected on the $(s, z)$-plane; no filtering needed because the fast and "noisy" variable $h$ is orthogonal to the projection plane. The equilibrium points and the attracting regions are indicated. The motion is seen to occur within a narrow layer between the equilibrium points. A $(s', h)$-plane, given by equation (22), is shown by a dashed line intersecting the layer of motion (note that the system's velocity approximately lies in the $(s', h)$-planes).

Figure 6b is the projection on the $(s, z)$-plane of the 222yr evolution path considered in Section 3.2. As can be seen, this path is confined to a relatively narrow layer in the phase space, as if the motion were occurring in the parallel planes stacked together. Equation (21a) explains this behavior. Given the parameter values reported in Table I, $c_2(s - s_*)$ and $\omega_y(e^z - b)$ are small in comparison to $c_1\dot{s}$, therefore it follows from equation (21a) that $|\dot{z} - c_1\dot{s}| \ll 1$. In geometrical terms, the expression $\dot{z} - c_1\dot{s}$ has the meaning of the velocity normal to the planes

$$z - c_1 s = C, \tag{22}$$

where $C$ is an integration constant. The planes defined by equation (22) are parallel to the plane containing the $h$-axis and the line $z = c_1 s$; we will henceforth, out of convenience, call these planes the $(s', h)$-planes, where $s'$ denotes the variable that measures distance along $z = c_1 s$ (Figure 6b).

Thus, the system's velocity approximately lies in the $(s', h)$-planes. Therefore, the amplitude of the planar movement dominates the path development, so that the system quickly traverses the relatively long distance in the $(s', h)$-planes between the attractors and moves slowly across the $(s', h)$-planes when it is captive to the attractors.

Figure 6 gives us first insights into the dynamics. The system crosses rapidly from the contraction regime attractor to the expansion regime attractor along a trajectory lying in one of the upper $(s', h)$-planes (Figure 6b). Once it is entrapped by the expansion attractor, the system switches to the dynamic of orbiting the attractor and slowly drifts toward it, while being acted upon by stochastic shocks $\xi_t$. Eventually, $\xi_t$ forces the system on a trajectory leading toward the contraction attractor in a lower $(s', h)$-plane (Figure 6b). Once entrapped there, the system begins its slow, spiraling ascent toward equilibrium until it is randomly thrown by $\xi_t$ onto a trajectory leading to the other attractor. And so the cycle repeats.

Figure 7a provides further details. It shows the evolution path in Figure 6a projected on the $(s'_m, h)$-plane given by equation (22) with $C = -1.00$ (this plane intersects the domain, to which the system's evolution is confined, roughly in the middle). The two entrapment regions are visible. Note that motion inside the entrapment regions is clockwise. Note further that the system tends to stay longer in the region where the economy is expanding than in that region where it is contracting. This asymmetry arises due to the positive long-term news component $\varepsilon$, which reflects technological progress.

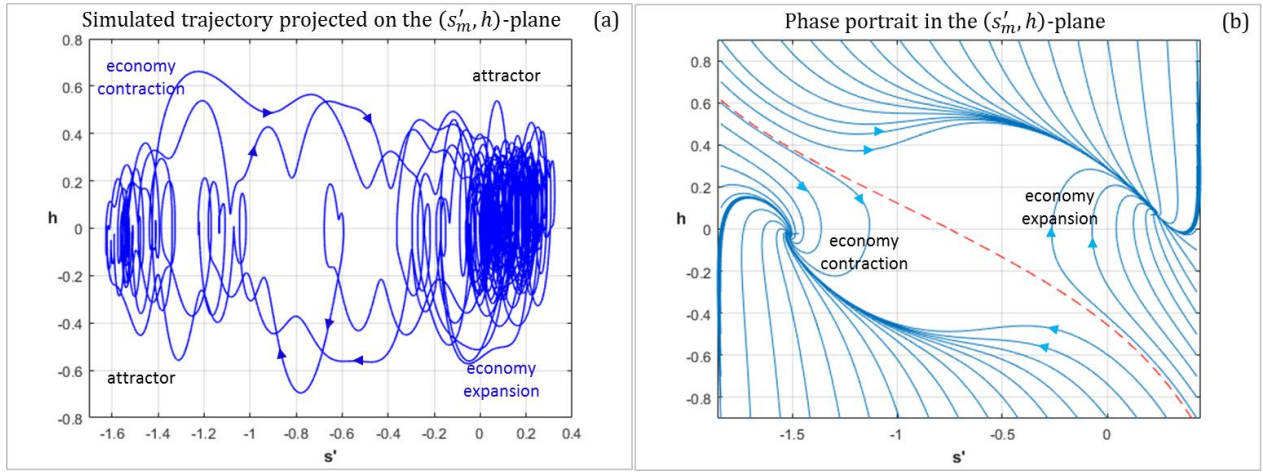

**Figure 7**: (a) The 14yr smoothed trajectory from Figure 6(a) projected on the $(s'_m, h)$-plane, given by equation (22) with $C = -1.00$, which intersects approximately in the middle of the layer of motion in Figure 6(b). (b) The phase portrait of dynamical system (21) ($\xi_t = 0$) shown in this same plane with the separatrix indicated by a dashed red line.

Because the motion is nearly planar, we can construct the system's phase portraits ($\xi_t = 0$) that lie approximately in different $(s', h)$-planes, given by equation (22) for different $C$. We will do this by emitting trajectories from the boundaries of the motion in these planes. This approximation holds reasonably well, except at the core of the attracting regions.

Figure 7b depicts the phase portrait in the $(s'_m, h)$-plane considered above. We observe that the two attractors are not connected – the system cannot cross the separatrix marked by the dashed red line. Therefore, $\xi_t \neq 0$ is necessary for the system to cross the separatrix and so escape the entrapment regions. When it is crossed, the system finds itself on a long trajectory that passes in the separatrix's vicinity. This trajectory swiftly takes the system to the other entrapment region where the system stays until $\xi_t$ randomly forces it across the separatrix and the system is carried back into the entrapment region whence it started.

The distance between each attractor and the separatrix represents the barrier that $\xi_t$ must overcome to thrust the system out of the entrapment regions and thus instigate a regime transition. It turns out that this distance is slowly fluctuating, so that the probability of escaping also gradually increases and decreases. To show this effect, we must examine in more detail the downwelling and upwelling motions that occur in the entrapment regions in the direction normal to the $(s', h)$-planes (equation (22) and e.g. Figure 7).

As has been discussed in Figure 6b, the system travels from the contraction attractor to the expansion attractor in the upper $(s', h)$-planes and it travels the opposite direction in the lower $(s', h)$-planes. Thus, the downwelling and upwelling motions take place between these planes. The phase portraits in such planes may allow us, therefore, to visualize the barrier as it is experienced by the system in the beginning and at the end of its ascent and descent.

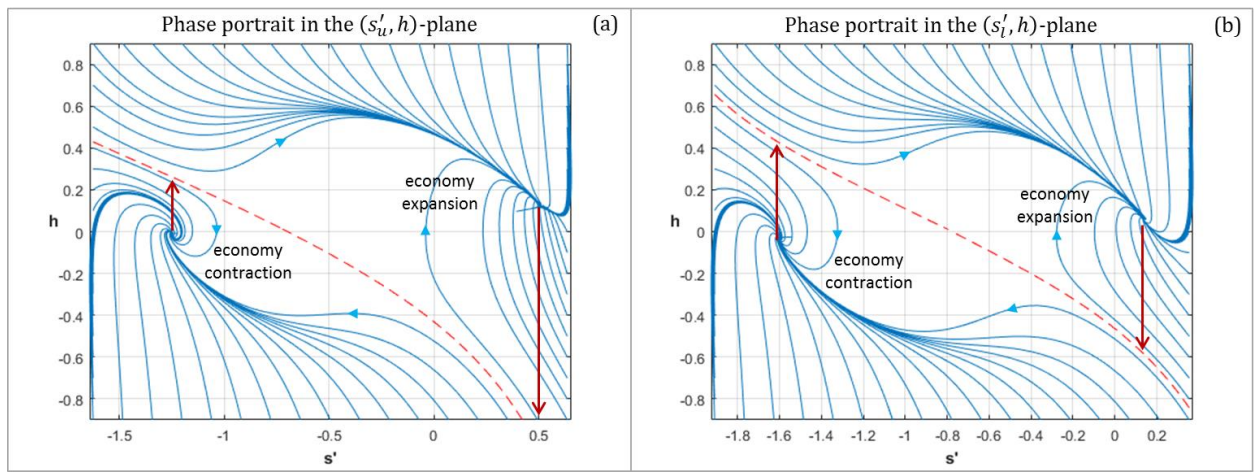

**Figure 8**: Two phase portraits of dynamical system (21) in (a) the plane $(s'_u, h)$ given by equation (22) for $C = -0.48$, located approximately at the top of the layer of motion (see Figure 6b); and (b) the plane $(s'_l, h)$ given by equation (22) for $C = -1.11$, located approximately at the bottom of the layer of motion (see Figure 6b). The red arrows show the distance in $h$ from the stable equilibrium

points to the separatrix and therefore indicate the magnitude of the barrier to the system's transition from one equilibrium to the other. In the expansion regime, the system descends from the upper plane of motion $(s'_u, h)$ toward the lower plane $(s'_l, h)$, so that the distance from the expansion equilibrium to the separatrix decreases, making it easier for $\xi_t$ – a news-driven stochastic shock in $h$ – to thrust the system across the separatrix. In the contraction regime, the system ascends from the lower plane of motion $(s'_l, h)$ toward the upper plane $(s'_u, h)$, the distance between the contraction equilibrium and the separatrix decreases, enabling $\xi_t$ to eventually make the system jump across the separatrix.

Figure 8a depicts the phase portrait in the upper plane $(s'_u, h)$ given by equation (22) for $C = -0.48$ and Figure 8b shows the phase portrait in the lower plane $(s'_l, h)$ given by equation (22) for $C = -1.11$. In the expansion regime, the barrier height in the upper $(s'_u, h)$-plane is greater than in the lower $(s'_l, h)$-plane. Thus, as the system begins its slow descent toward equilibrium, the probability that it jumps across the separatrix, ending expansion prematurely, is relatively low. This probability increases as the system descends further toward the $(s'_l, h)$-plane, gradually opening a window of opportunity for an adverse news event to randomly instigate regime transition. Similarly, in the contraction regime, the probability of regime transition is increasing as the system ascends from the $(s'_l, h)$-plane toward the $(s'_u, h)$-plane.

The above-described mechanism underlying business cycles can be classified as coherence resonance, a recently studied phenomenon whereby noise applied to a dynamical system leads to a quasiperiodic response (Pikovsky and Kurths, 1997). Coherence resonance takes its classic form where the dynamical system is tuned in a subcritical regime close to the emergence of a limit cycle. In this situation, the system's phase portrait may contain the unclosed largescale trajectories in the vicinity of the about-to-emerge periodic limit cycle. A certain amount of noise added to the system will "patch up" these trajectories creating a limit cycle which is quasiperiodic. The phase portrait studied in this paper differs from the classic case because there are two attractors at which the trajectories end (Figures 7b and 8). Still, the mechanism is essentially the same as the noise forces the system across the separatrix – effectively reconnecting the trajectories between the attractors and inducing quasiperiodic behaviors.[23, 24]

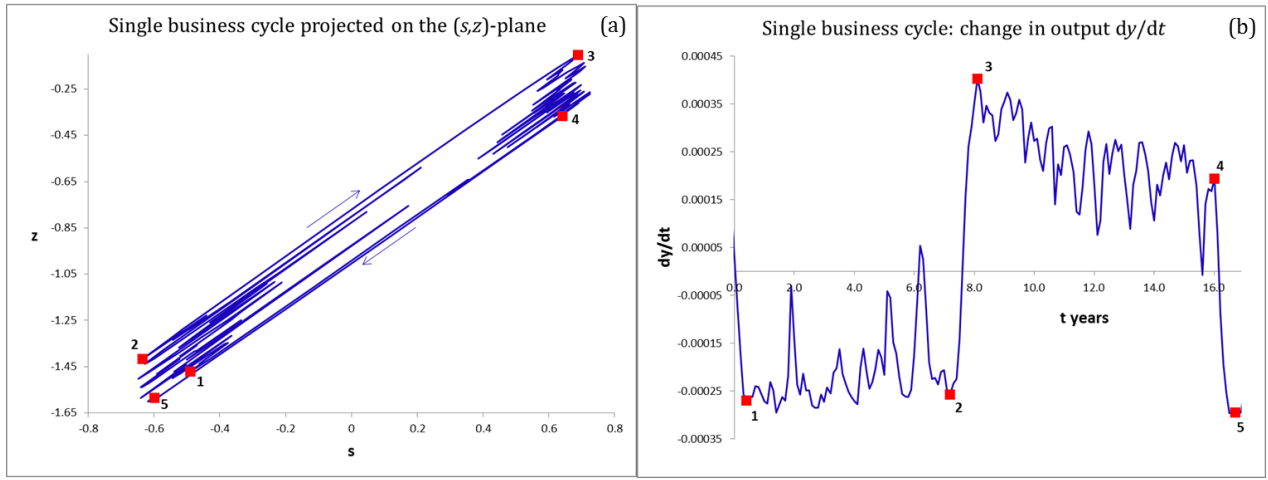

**Figure 9**: A single business cycle (a) projected on the $(s, z)$-plane and (b) plotted as the evolution of output growth with time, $\dot{y}(t)$. The characteristic regimes are separated by the red squares.

---

[23] Coherence resonance emerges in various physical, chemical and biological systems. In particular, it has been studied in connection with nerve cell dynamics. See a review by Lindner et al. (2004).

[24] We take note of Beaudry et al. (2017) that develops a DSGE model with herding. This model produces quasiperiodic fluctuations explained as a limit cycle stochastically perturbed by productivity shocks. As such, noise acts there to detune an existing limit cycle, in contrast with the coherence resonance where noise patches up an incomplete "limit cycle".

Finally, we would like to connect the system's behavior in the phase space to the real-world description. To do this, let us consider a single simulated business cycle both as projected on the ($s, z$)-plane (Figure 9a) and as a function of time (Figure 9b). The cycle's duration is 17 years. In interval 1-2 the economy undergoes contraction because the rate of change in output is negative (Figure 9b). This interval corresponds to the state of entrapment by the contraction attractor in the phase space (Figure 9a). Interval 2-3 describes a rapid transition from contraction to expansion, which coincides in the phase space with the escape from the contraction attractor and the entrapment by the expansion attractor. During interval 3-4, the economy's output is growing since the economy-market system is trapped by the expansion attractor. Note how this growth slows (Figure 9b) as the coupled system slowly drifts toward the attractor (Figure 9a), setting the stage for regime transition. Such a transition occurs during interval 4-5: the economy commences contraction as the result of having escaped the expansion attractor and being captured by the contraction attractor.